# Inverse melting of the vortex lattice

Nurit Avraham[1], Boris Khaykovich[1*], Yuri Myasoedov[1], Michael Rappaport[1], Hadas Shtrikman[1], Dima E. Feldman[1,2], Tsuyoshi Tamegai[3], Peter H. Kes[4], Ming Li[4], Marcin Konczykowski[5], Kees van der Beek[5], and Eli Zeldov[1]

[1]*Department of Condensed Matter Physics, The Weizmann Institute of Science, Rehovot 76100, Israel*

[2]*Landau Institute for Theoretical Physics, 142432 Chernogolovka, Moscow region, Russia*

[3]*Department of Applied Physics, The University of Tokyo, Hongo, Bunkyo-ku, Tokyo 113-8656, and CREST, Japan Science and Technology Corporation (JST), Japan*

[4]*Kamerlingh Onnes Laboratory, Leiden University, P.O. Box 9504, 2300 RA Leiden, The Netherlands*

[5]*CNRS, UMR 7642, Laboratoire des Solides Irradies, Ecole Polytechnique, 91128 Palaiseau, France*

*Present address: Department of Physics, Massachusetts Institute of Technology, Cambridge, MA 02139

**Inverse melting, in which a crystal reversibly transforms into a liquid or amorphous phase upon decreasing the temperature, is considered to be very rare in nature[1]. The search for such an unusual equilibrium phenomenon is often hampered by the formation of nonequilibrium states which conceal the thermodynamic phase transition, or by intermediate phases, as was recently shown in a polymeric system[2]. Here we report a first-order inverse melting of the magnetic flux line lattice in $Bi_2Sr_2CaCu_2O_8$ superconductor. At low temperatures, the material disorder causes significant pinning of the vortices, which prevents observation of their equilibrium properties. Using a newly introduced 'vortex**

**dithering' technique[3] we were able to equilibrate the vortex lattice. As a result, direct thermodynamic evidence of inverse melting transition is found, at which a disordered vortex phase transforms into an ordered lattice with increasing temperature. Paradoxically, the structurally ordered lattice has larger entropy than the disordered phase. This finding shows that the destruction of the ordered vortex lattice occurs along a unified first-order transition line that gradually changes its character from thermally-induced melting at high temperatures to a disorder-induced transition at low temperatures.**

Local magnetization measurements, using microscopic Hall sensors[4], were carried out on a number of high quality optimally-doped $Bi_2Sr_2CaCu_2O_8$ (BSCCO) crystals, grown in two laboratories[5,6]. A dc magnetic field $H_a$ was applied parallel to the crystalline c-axis, while an ac transverse field $H_{ac\perp}$ was applied along the ab-planes (see caption of Fig. 1 for details).

Willemin *et al.* have recently shown[3] that in $YBa_2Cu_3O_7$ at temperatures close to $T_c$, 'vortex dithering' by a transverse ac field reduces the irreversible magnetization caused by vortex pinning. We find that in BSCCO crystals this method can fully suppress the magnetic hysteresis even at low temperatures, down to about 30 K, as shown in Fig. 1a. The Abrikosov vortices in BSCCO can be regarded as a stack of Josephson-coupled pancake vortices[7] in the individual $CuO_2$ planes. The in-plane field $H_{ac\perp}$ readily penetrates through the sample[8] in the form of Josephson vortices (JV) residing in-between the $CuO_2$ planes[9]. The main effect, which is of interest here, is that a pancake vortex, located in a $CuO_2$ plane immediately above a JV, is displaced a small distance along the direction of the JV relative to the neighboring pancake vortex residing one $CuO_2$ layer underneath[9]. For small values of $H_{ac\perp}$ the probability of such a close proximity between a pancake and a JV is low. Thus, in our sample geometry, an average pancake experiences only a few such periodic 'intersections' during one ac cycle with a duration of each intersection of less than 1% of the ac period. $H_{ac\perp}$ therefore induces a weak local ac agitation of pancakes, which assists thermal activation in relaxing the irreversible magnetization and in approaching thermal equilibrium.

In addition to bulk pinning, a substantial part of the magnetic hysteresis in BSCCO crystals is caused by surface[10] and geometrical barriers[11]. $H_{ac\perp}$ also efficiently suppresses this source of hysteresis. The JV, which periodically cross the sample edges, instantaneously 'slice' the Abrikosov vortices into short segments or individual pancakes, thus apparently facilitating[10] vortex penetration though the surface and geometrical barriers. At lower temperatures the equilibration of vortices by $H_{ac\perp}$ becomes progressively more difficult, as shown in Fig. 1b at $T = 26$ K, where our maximum $H_{ac\perp}$ is not sufficient to completely eliminate the hysteresis.

We now focus on the main topic of vortex matter phase transitions. At elevated temperatures the vortex lattice undergoes a first-order melting transition[4,8,12,13]. The melting line $B_m(T)$ was found to apparently terminate at some critical point $T_{cp}$, below which a "second magnetization peak" line $B_{sp}(T)$ emerges[14,15,16,17]. In BSCCO crystals the $B_{sp}(T)$ line is rather horizontal on the $B$-$T$ phase diagram. On crossing this line by a field sweep, a pronounced "second peak" is observed in magnetization loops[14] (see Fig. 1), which reflects a transformation of a quasi-ordered lattice into a disordered amorphous state with enhanced vortex pinning.

The first-order nature of the melting transition $B_m(T)$ is manifested by a small step in equilibrium magnetization, as described in Fig. 2. At high temperatures, $H_{ac\perp}$ has no observable effect on this step. As the temperature is decreased along $B_m(T)$ towards $T_{cp}$, magnetic hysteresis starts to set in gradually, as seen for example in Fig. 2a. Here, at $T > T_{cp}$, the magnetization step is still clearly resolved, but it is partially suppressed by the hysteresis in the vortex solid phase below the transition. Application of $H_{ac\perp}$ fully removes the hysteresis and enhances the magnetization step. $H_{ac\perp}$ also results in a small broadening of the transition. We attribute this broadening to the fact that a transverse field slightly decreases[9] the melting field $B_m$. The small variation of $B_m$ during the ac cycle of $H_{ac\perp}$ should therefore cause some smearing of the magnetization step.

Figure 2b shows the behaviour at $T = 38$ K, just below $T_{cp}$, where the first-order transition was believed to be absent. At this temperature significant hysteresis is observed and the second magnetization peak starts to develop. Remarkably, application of $H_{ac\perp}$ eliminates the hysteresis and instead a step in reversible magnetization emerges, which reveals the existence of a first-order transition. A similar phenomenon is found at lower temperatures, as shown in Fig. 2c, where a clear second magnetization peak turns into a step in magnetization upon application of $H_{ac\perp}$ (see also Fig. 1a). This finding leads to a number of important conclusions, including: (i) the first-order transition does not terminate at $T_{cp}$, (ii) the vortex matter displays inverse melting, (iii) the underlying mechanism of the first-order transition is different above and below $T_{cp}$, and (iv) the disorder-driven transition at the second magnetization peak is of first order.

(i) The first conclusion is that the previously reported termination of the first-order transition[4] at $T_{cp}$ does not reflect a real critical point, but rather an experimental limitation. Below $T_{cp}$ the dynamics becomes too slow to achieve vortex equilibration on typical experimental time scales, thus preventing observation of the equilibrium magnetization step. $H_{ac\perp}$ accelerates the equilibration process[3] and thus facilitates the detection of the magnetization step down to significantly lower temperatures.

(ii) Observation of the magnetization step below $T_{cp}$ is not just an extension of the range of the first-order transition, but rather provides qualitatively new information. Figure 3 shows the location of the transition line on the $B$-$T$ phase diagram. Without $H_{ac\perp}$, the first-order transition is observed only at temperatures above $T_{cp}$, where $B_m(T)$ has a negative slope $dB_m/dT$. Interestingly, $B_m(T)$ becomes horizontal[4] on approaching $T_{cp}$. By applying $H_{ac\perp}$ we find that the first-order transition extends to significantly lower temperatures and shows an inverted behaviour of a positive slope below $T_{cp}$. Hence, in this region the vortex matter displays the very rare phenomenon of inverse melting in which an ordered lattice (Bragg glass[18,19]) reversibly transforms into a liquid or amorphous phase upon cooling.

(iii) In a thermally-driven vortex melting transition, the liquid phase has to be on the high-temperature side of the transition since thermal fluctuations always increase with temperature. The observed inverse melting behaviour therefore implies that the first-order transition below $T_{cp}$ must have a different underlying nature, which is disorder-driven rather than thermally-driven. The main conceptual difference is that in thermal melting the decrease in elastic energy of the lattice at the transition is compensated by a gain in entropy of the liquid. In a disorder-driven transition, in contrast, the loss of elastic energy is balanced by a gain in pinning energy, since the 'wiggling' entangled vortices adapt more efficiently to the pinning landscape induced by the quenched material disorder[20]. At low fields, the elastic energy of the lattice is higher than the pinning energy, giving rise to an ordered state[18,19,20]. With increasing field the elastic energy decreases relative to pinning, leading to a structural transformation into a disordered phase when the two energies become comparable. Our finding shows that this non-thermal transition is of first order. At low temperatures the shape of the disorder-driven transition line should be nearly temperature independent or may show downward curvature due to the temperature dependence of the microscopic parameters[21]. At intermediate temperatures, however, even though the transition remains disorder-driven, thermal smearing of the pinning potential progressively reduces the pinning energy, leading to an upturn in the transition line[20]. This effect of pinning reduction is apparently at the origin of the observed inverse melting behaviour. An upturn in the shape of the second magnetization peak line has been noted previously[5,14]; however, the results presented here are the first evidence of a thermodynamic inverse melting transition of the vortex lattice.

(iv) The extended $B_m(T)$ transition line coincides with the location of the $B_{sp}(T)$ line at lower temperatures, as seen in the inset of Fig. 3, which demonstrates that the two phenomena have a common origin. The second magnetization peak is the dynamic characteristic of the transition, reflecting the enhanced vortex pinning and the viscous nature of the disordered amorphous phase, whereas the reversible magnetization step is its thermodynamic signature. The revealed first-order nature of the second peak is consistent with recent conclusions based on Josephson plasma resonance studies[22] and transient measurements[23,24], as well as several numerical simulations[25,26,27]. Thus, the breakdown of the Bragg glass is apparently always a first-order transition that occurs either through a thermal, disorder-driven, or combined mechanism. While thermally-driven melting is possibly restricted to high-temperature superconductors, the disorder-driven transition should be a more general phenomenon and is apparently at the heart of the ubiquitous peak effect in low-$T_c$ superconductors[28].

Figure 4 shows the magnetization step $\Delta B$ and the corresponding entropy change $\Delta S = -(dH_m/dT)\Delta B/4\pi$ in the vicinity of $T_{cp}$. $\Delta B$ is approximately constant in this region. The entropy change $\Delta S$ vanishes at $T_{cp}$ because $dH_m/dT = 0$, and becomes negative below $T_{cp}$. A negative $\Delta S$ means that, paradoxically, the ordered lattice has larger entropy than the disordered phase. Since the ordered lattice has no dislocations and is structurally more ordered, the extra entropy must arise from additional degrees of freedom. In the amorphous phase the lattice structure is broken and the vortices wander out of their unit cells and become entangled; yet, their thermal fluctuations on short time scales are small due to enhanced pinning. In the ordered lattice, on the other hand, there is no entanglement and no large scale vortex wandering; however, thermal fluctuations within the unit cell are larger, apparently resulting in larger entropy. $\Delta S < 0$ also means that the latent heat $T_m \Delta S$ is negative, and hence the sample releases heat upon melting on increasing field, in contrast to the usual melting in which the sample absorbs heat. Note, however, that by crossing the transition lines by increasing the temperature, the latent heat is always positive, as dictated by thermodynamics, and the high temperature phase always has larger entropy. The observation of a rather constant **DB** in Fig. 4 resolves another previously reported[4] inconsistency in the behaviour near $T_{cp}$: At a true critical point **DB** is expected to vanish continuously, whereas experimentally a rather constant **DB** was found to disappear abruptly at $T_{cp}$. The consistent values of **DB** above and below $T_{cp}$ clearly indicate a smooth continuation of the first-order transition and the absence of a real critical point. We emphasize, however, that the vortex system may not be in full thermal equilibrium in the presence of $H_{ac\perp}$ agitation. Therefore the relatively large values of **DB** in Fig. 4, which are found to be sample dependent, should be further investigated.

Finally, an important open question is whether the two disordered states of the vortex matter, liquid and amorphous, represent different thermodynamic phases or just a continuous slowdown of vortex dynamics. Recent numerical calculations lead to contradictory conclusions[25,26,27,29]. We do not observe any sharp features along the depinning line $T_d$, which usually extends upward[30] from $T_{cp}$ and is believed to separate the two phases. A strongly first-order $T_d$ transition would require a sharp change in slope of $B_m(T)$ at $T_{cp}$, which we do not observe. Therefore, the $T_d$ line in BSCCO is unlikely to be a first-order transition; it could be either a continuous transition or a dynamic crossover. Similarly, in $YBa_2Cu_3O_7$ the structure of the phase diagram is controversial, and it is as yet unclear whether the second magnetization peak line merges with the melting line at the critical point[16] or at a lower field.

1. Greer, A. L. Too hot to melt. *Nature* **404**, 134-135 (2000).

2. Rastogi, S., Höhne, G. W. H. and Keller, A. Unusual pressure-induced phase behaviour in crystalline Poly(4-methylpentence-1): Calorimetric and spectroscopic results and further implications. *Macromolecules* **32**, 8897-8909 (1999).

3. Willemin, M. *et al.* First-order vortex-lattice melting transition in $YBa_2Cu_3O_7$ near the critical temperature detected by magnetic torque. *Phys. Rev. Lett.* **81**, 4236-4239 (1998).

4. Zeldov, E. *et al.* Thermodynamic observation of first-order vortex-lattice melting transition. *Nature* **375**, 373-376 (1995).

5. Ooi, S., Shibauchi, T. and Tamegai, T. Evolution of vortex phase diagram with oxygen-doping in $Bi_2Sr_2CaCu_2O_{8+y}$ single crystals. *Physica C* **302**, 339-345 (1998).

6. T. W. Li *et al*. Growth of $Bi_2Sr_2CaCu_2O_{8+x}$ single-crystals at different oxygen ambient pressures. *J. Cryst. Growth* **135**, 481-486 (1994).

7. Blatter, G. *et al*. Vortices in high-temperature superconductors. *Rev. Mod. Phys.* **66**, 1125-1388 (1994).

8. Pastoriza, H., Goffman, M. F., Arribere, A. and de la Cruz, F. First order phase transition at the irreversibility line of $Bi_2Sr_2CaCu_2O_8$. *Phys. Rev. Lett.* **72**, 2951-2954 (1994).

9. Koshelev, A. E. Crossing lattices, vortex chains, and angular dependence of melting line in layered superconductors. *Phys. Rev. Lett.* **83**, 187-190 (1999).

10. Burlachkov, L., Koshelev, A. E. and Vinokur, V. M. Transport properties of high-temperature superconductors: surface vs bulk effect. *Phys. Rev. B* **54** 6750-6757 (1996).

11. Zeldov, E. *et al.* Geometrical barriers in high-temperature superconductors. *Phys. Rev. Lett.* **73**, 1428-1431 (1994).

12. Safar, H. *et al*. Experimental evidence for a first-order vortex-lattice-melting transition in untwinned single crystal $YBa_2Cu_3O_7$. *Phys. Rev. Lett.* **69**, 824-827 (1992).

13. Kwok, W. K. *et al*. Vortex lattice melting in untwinned and twinned single-crystals of $YBa_2Cu_3O_{7-\delta}$. *Phys. Rev. Lett.* **69**, 3370-3373 (1992).

14. Khaykovich, B. *et al.* Vortex lattice phase transitions in $Bi_2Sr_2CaCu_2O_8$ crystals with different oxygen stoichiometry. *Phys. Rev. Lett*. **76**, 2555-2558 (1996).

15. Chikumoto, N., Konczykowski, M., Motohira, N., and Malozemoff, A. P. Flux-creep crossover and relaxation over surface barriers in $Bi_2Sr_2CaCu_2O_8$ crystals. *Phys. Rev. Lett.* **69**, 1260-1263 (1992).

16. Deligiannis, K. *et al.* New features in the vortex phase diagram of $YBa_2Cu_3O_7$. *Phys. Rev. Lett*. **79**, 2121-2124 (1997).

17. Cubitt, R. *et al.* Direct observation of magnetic flux lattice melting and decomposition in the high-$T_c$ superconductor $Bi_2Sr_2CaCu_2O_8$. *Nature* **365**, 407-411 (1993).

18. Giamarchi, T. and Le Doussal, P. Elastic theory of pinned flux lattice. *Phys. Rev. Lett*. **72**, 1530-1533 (1994).

19. Nattermann, T. and Scheidl, S. Vortex-glass phases in type-II superconductors. *Adv. Phys.* **49**, 607-704 (2000).

20. Ertas, D. and Nelson, D. R. Irreversibility, entanglement and thermal melting in superconducting vortex crystals with point impurities. *Physica C* **272**, 79-86 (1996).

21. Giller, D. *et al*. Disorder-induced transition to entangled vortex-solid in Nd-Ce-Cu-O crystal. *Phys. Rev. Lett.* **79**, 2542-2545 (1997).

22. Gaifullin, M. B. *et al*. Abrupt change of Josephson plasma frequency at the phase boundary of the Bragg glass in $Bi_2Sr_2CaCu_2O_{8+d}$. *Phys. Rev. Lett.* **84**, 2945-2948 (2000).

23. van der Beek, C. J., Colson, S., Indenbom, M. V. and Konczykowski, M. Supercooling of the disordered vortex lattice in $Bi_2Sr_2CaCu_2O_{8+d}$. *Phys. Rev. Lett.* **84**, 4196-4199 (2000).

24. Giller, D., Shaulov, A., Tamegai, T. and Yeshurun, Y. Transient vortex states in $Bi_2Sr_2CaCu_2O_{8+d}$ crystals. *Phys. Rev. Lett.* **84**, 3698-3701 (2000).

25. Kierfeld, J. and Vinokur, V. Dislocations and the critical endpoint of the melting line of vortex line lattices. *Phys. Rev. B* **61**, R14928-R14931 (2000).

26. Nonomura, Y. and Hu, X. Effects of point defects on the phase diagram of vortex states in high-$T_c$ superconductors in B||c axis. Preprint cond-mat/0011349.

27. Olsson, P. and Teitel, S. Disorder driven melting of the vortex line lattice. Preprint cond-mat/0012184.

28. Paltiel, Y. *et al*. Instabilities and disorder-driven first-order transition of the vortex lattice. *Phys. Rev. Lett.* **85**, 3712-3715 (2000).

29. Reichhardt, C., van Otterlo, A., and Zimányi, G. T. Vortices freeze like window glass: the vortex molasses scenario. *Phys. Rev. Lett.* **84**, 1994-1997 (2000).

30. Fuchs, D. T. *et al*. Possible new vortex matter phases in $Bi_2Sr_2CaCu_2O_8$. *Phys. Rev. Lett.* **80**, 4971-4974 (1998).

Acknowledgements: We thank V. B. Geshkenbein for valuable discussions. This work was supported by the Israel Science Foundation - Center of Excellence Program, by Minerva Foundation, Germany, by the Mitchell Research Fund, and by the Grant-in-Aid for Scientific Research from the Ministry of Education, Science, Sports and Culture, Japan. DEF acknowledges the support by the Koshland Fellowship and RFBR grant. PK and ML acknowledge support of the Dutch Foundation FOM. EZ acknowledges the support by the Fundacion Antorchas - WIS program and by the Ministry of Science, Israel.

Correspondence and requests for materials should be addressed to N.A. (e-mail: nurit.avraham@weizmann.ac.il).

Figure 1. Local magnetization loops in BSCCO crystals with and without (o) 'vortex dithering'. A number of optimally doped BSCCO crystals with $T_c \cong 90$ K were investigated. We present here results on three crystals: A - $160 \times 600 \times 25$ μm$^3$, B - $170 \times 600 \times 20$ μm$^3$, and C - $190 \times 900 \times 30$ μm$^3$. The samples were mounted on an array of $10 \times 10$ μm$^2$ GaAs/AlGaAs Hall sensors. A dc magnetic field $H_a$ was applied perpendicular to the surface of the sensors (parallel to the crystalline c-axis). A transverse ac field $H_{ac\perp}$ (1 kHz with amplitude of up to 80 Oe), applied parallel to the ab-planes of the crystals, equilibrates the vortices by 'dithering' and suppresses the magnetic hysteresis. The Hall sensors are insensitive to the transverse field and measure only the dc c-axis component $B_z$ of the local induction. (a) Lowest temperature, 30 K, at which fully reversible magnetization in crystal A is attained for our maximum $H_{ac\perp}$ of about 80 Oe. The dc magnetization loop displays the second magnetization peak above 300 Oe, which is due to the enhanced vortex pinning in the disordered phase. Application of $H_{ac\perp}$ results in a reversible magnetization curve which reveals a small step in magnetization (barely seen on this scale) at the original location of the second magnetization peak. (b) Example of the local magnetization loop at a lower temperature (26 K) in crystal B. The dc loop shows large hysteresis and a pronounced second magnetization peak. The applied $H_{ac\perp}$ is not sufficient to obtain reversible magnetization. The 'dithering' efficiency of $H_{ac\perp}$ is significantly different in the two vortex phases: For the same width of the original dc loop, much stronger suppression of the hysteresis is observed in the ordered phase as compared to the disordered phase above the second magnetization peak.

Figure 2. Reversible magnetization steps revealed in BSCCO crystals by 'vortex dithering' at various temperatures. At high temperatures the first-order

magnetization step occurs in the region of reversible magnetization. In this case $H_{ac\perp}$ has no effect on the magnetization step up to our maximum transverse field of 80 Oe. Also there are no heating effects, as confirmed by the absence of a temperature shift in $B_m(T)$. (a) At intermediate temperatures, such as $T = 44$ K $> T_{cp} \cong 40$ K (crystal C), hysteretic magnetization is observed below the transition (o). Application of $H_{ac\perp}$ removes the residual hysteresis and enhances the magnetization step. Here the curve with $H_{ac\perp}$ was displaced downward for clarity. (b) At temperatures slightly below $T_{cp}$ significant hysteresis develops in the dc magnetization loops and the beginning of the second-magnetization-peak behaviour is observed in the form of a pronounced change in the width of the loop. $H_{ac\perp}$ results in reversible magnetization and reveals a clear magnetization step that appears instead of the second magnetization peak (crystal C, $T = 38$ K). (c) At still lower temperatures a fully developed second magnetization peak is observed in the dc magnetization loop which is replaced by a step in the reversible magnetization in the presence of $H_{ac\perp}$ (crystal B, $T = 32$ K).

Figure 3. The first-order transition line and the inverse melting obtained with 'vortex dithering'. Inset: The first-order $B_m(T)$ line (●) along with the second magnetization peak line $B_{sp}(T)$ (o) over a wide temperature range in BSCCO crystal B. $B_{sp}(T)$ was measured without $H_{ac\perp}$ and the error bars reflect the uncertainty in the determination of the location of the $B_{sp}$ transition due to its larger width as compared to the width of the first-order magnetization step. Main panel: An expanded view of the first-order transition line in the vicinity of $T_{cp}$. The first-order transition was observed at all temperatures at which fully reversible magnetization was achieved by 'vortex dithering'. At $T < 31$ K no reversible magnetization could be attained in this crystal with our maximum $H_{ac\perp}$ and hence no magnetization step could be revealed.

Figure 4. Magnetization step and the negative latent heat. The top panel is the height of the magnetization step $\Delta B$ in BSCCO crystal B in the vicinity of $T_{cp}$ obtained in the presence of $H_{ac\perp}$. Bottom panel shows the corresponding calculated entropy change $\Delta S$ (or the latent heat $T\Delta S$). The negative $\Delta S$ below $T_{cp}$ is the region of the inverse melting transition of the vortex lattice. In this region on increasing field the melting of the lattice is accompanied by a negative latent heat. On crossing this transition line by increasing the temperature, however, the opposite process of crystallization of the amorphous phase occurs, which is accompanied by a positive latent heat.

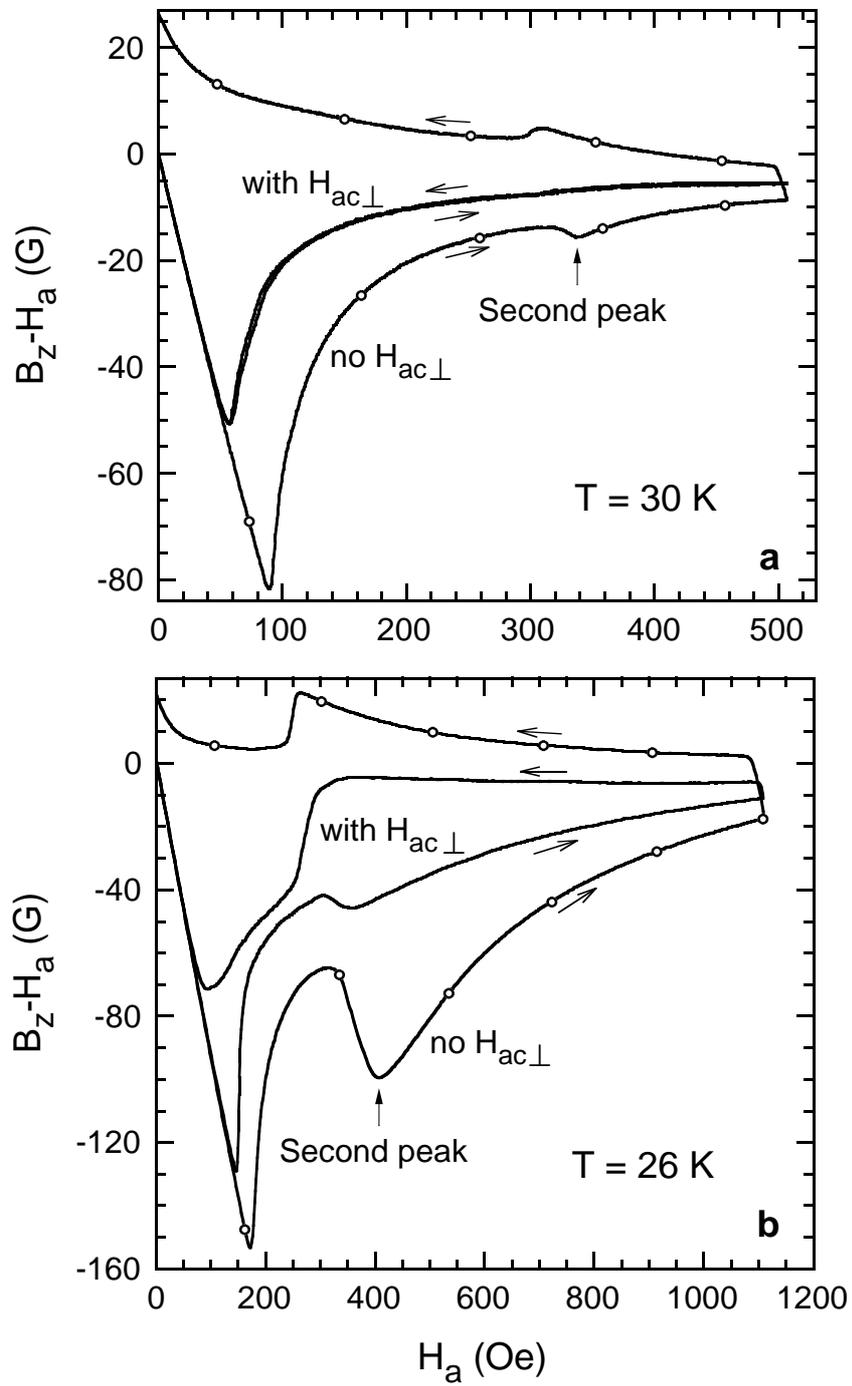

Fig. 1, Avraham et al.

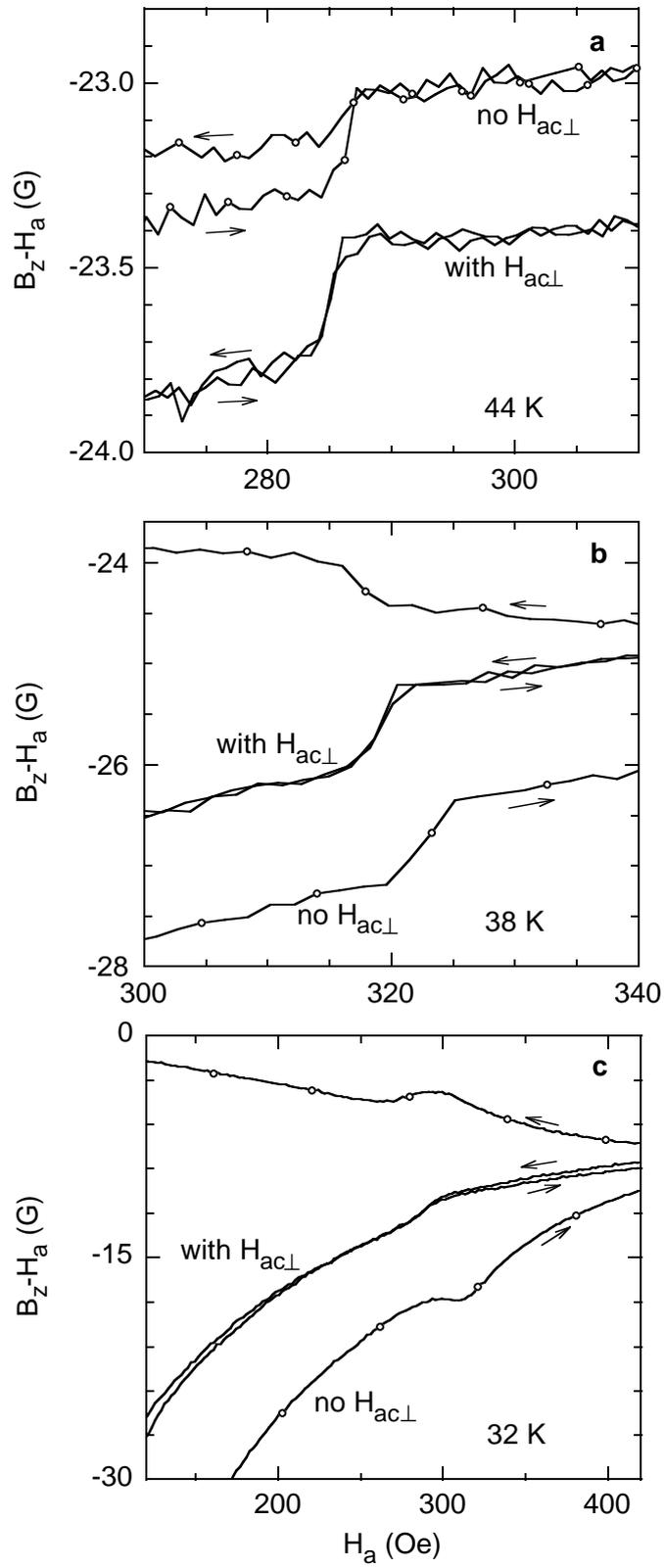

Fig. 2, Avraham et al.

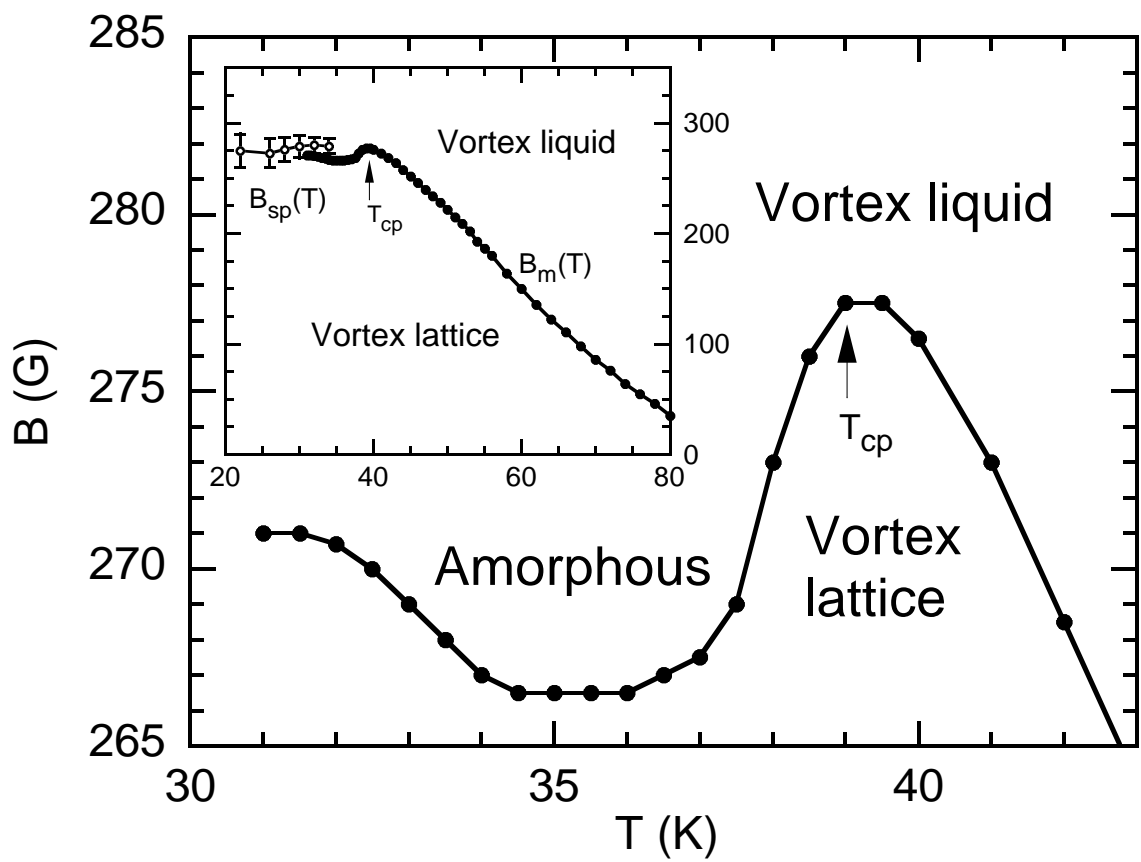

Fig. 3, Avraham et al.

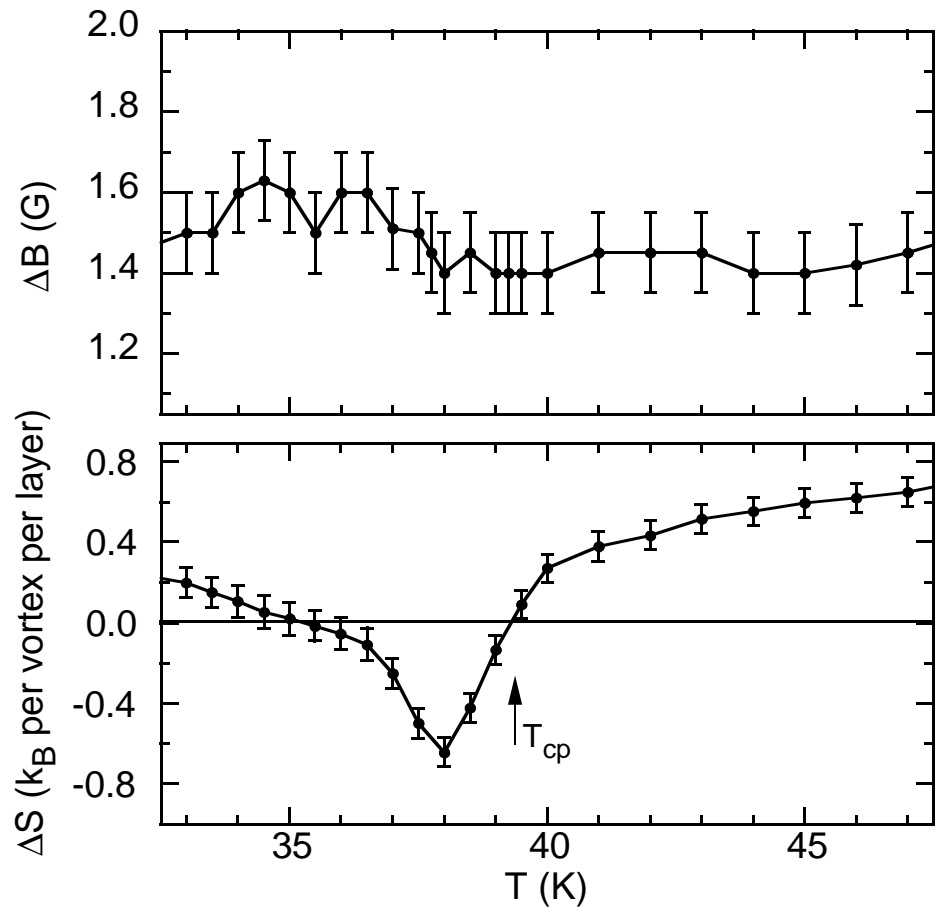

Fig. 4, Avraham et al.